\documentstyle[12pt]{article}

\begin{document}
\begin{flushright}
{BI-TP 94/15}
\end{flushright}
\begin{flushright}
{May 1994}
\end{flushright}
\begin{center}
{\bf {Energy Losses (Gains) of Massive Coloured Particles
in Stochastic Colour Medium}}
\end{center}
\begin{center}
{{\bf{Andrei Leonidov}}\footnote{Alexander von Humboldt Fellow}}
\end{center}
\begin{center}
{\it Physics Department\\ University of Bielefeld 33615
Bielefeld,Germany\\ and \\Theoretical Physics Department\\ P.N.Lebedev
Physics Institute, 117924 Leninsky pr.53, Moscow, Russia}\\
\end{center}
\medskip
\begin{abstract}
The propagation of massive coloured particles
 in stochastic background chromoelectric
field is studied using the semiclassical equations of motion.
Depending on the nature of the stochastic  background
 we obtain the formulae for  the energy losses
of heavy coloured projectile in nonperturbative hadronic medium and for
the energy gains in the stochastic field present, e.g., in the turbulent
plasma. The result appears to be significantly
dependent on the form of the correlation function of
stochastic external field.
\end{abstract}

\newpage
{\bf{1.}} It has become a part of the standard QCD wisdom that the
physics of heavy quarks provides unique possibilities for studying
both perturbative and non-perturbative phenomena. From the point of view
of perturbation theory, due to the asymptotic freedom the processes
involving heavy quarks tend to evolve at small distances [1]. At the same
time the heavy quarkonia are a unique testing device for  studying the
nonperturbative effects. One of the best examples
is the appearence of the vacuum condensates in the
description of heavy quarkonia physics in the framework of sum
rules method [2].

Of a particular interest to high energy physics are the unique abilities
of heavy quarkonia to probe the properties of a complicated medium
arising in the high energy nucleus-nucleus and hadron-nucleus
collisions.  In particular charmonium, when propagating through the
quark-gluon plasma, effectively measures its temperature giving a
possibility of its determination through the $J/\Psi$ to $J/\Psi'$ ratio
[3]. Another sensitive point, which was a subject of recent discussions,
is a drastic difference between the interaction of heavy quarkonium with
the medium, when the heavy quarkonium is in the singlet and octet colour
states respectively (see e.g.[4]).
In particular, such effects can be very
important for the description of the hard region of spectra in the
longitudinal momenta of charmonium, produced in hadron-nucleus
collisions [5]. Roughly speaking, the small dipole interaction
 of the singlet
projectile with the coloured medium is supposed to be responsible for
colour transparency, and that of an octet one for colour opacity.
Let us note, that the production of an initial octet then emitting a
gluon and becoming a singlet constitutes a part of the perturbative
contribution to the production of the heavy vector mesons [6].

Another situation of considerable interest to the high energy physics is
a propagation of a heavy quark through the hadronic (nuclear) medium. As
the process of creation of heavy quarks can often be reliably calculated
in perturbative QCD, one hopes to get an information on the medium
through which the heavy quark (heavy quark jet) propagates.

One of the models for the interior of the hadronic medium
is that of the soft stochastic colour fields giving rise to the
phenomenologically introduced condensates. Different variants of
this model were successfully
applied both for the explanation of the origin of confinement [7] and for
more practical calculations such as those of the meson spectra (see e.g.
[8,9])
and of the soft hadron-hadron scattering cross-sections
[10]. Below we shall consider the application
of this language to a case of a massive colour projectile in a stochastic
chromoelectric field.
In this situation it is interesting to know, how the
interaction with the medium changes the properties of the incident
particle
(for example, its energy, etc.).  We shall consider the changes of
the projectile energy due to the interaction with the stochastic
chromoelectric external field. It will be shown that the particle can either
lose or gain energy depending on the model adopted for the description
of the hadronic medium. The mechanism of this stochastic decceleration
(acceleration) is completely analogous to the phenomenon of stochastic
plasma heating
well-known in plasma physics (see, e.g., [11]). In this case when the
stochastic electric fields  are present in the plasma (for example, this
is a case for the turbulent plasma), the charged particles are
accelerated by it. This model was succsefully applied
to a description of the cosmic rays spectrum, where the charged
particles are accelerated by a turbulent cosmic plasma (see, e.g,
[12]).

{\bf{2.}}  Let us consider the propagation of a particle having a
non-abelian charge in some representation of a gauge group in the
external non-abelian field. The cooresponding quasiclassical
equations of motion are well-known [13,14] and read
\begin{equation}
\left \{
\begin{array}{lcl}
m{\dot{u}}_{\mu} &=& g Q^a(\tau) G^a_{\mu \nu}
 (x(\tau)){\dot{x}}^{\nu} \nonumber \\
{\dot{Q}}_a (\tau) &=& -g f_{abc}{\dot{x}}^{\mu} (\tau) A^b_{\mu}(x(\tau))
Q^c(\tau)
\end{array}
\right .
\end{equation}
where the derivative in the left hand side is taken with respect to a
proper time $\tau$, $m$ is a mass of an incident particle,
$g$ is a colour charge, $Q_a$ are the components of a colour spin,
$f_{abc}$ are the structure constants of a corresponding
Lie algebra, $x_{\mu} (\tau)$ is the particle trajectory in the
space-time, $u^{\mu}(\tau)$ is a corresponding four-velocity, $A^b_{\mu}$ are
the external gauge field potentials
forming a stochastic ensemble and $G^a_{\mu \nu}$ is a corresponding
field strength.  The first equation describes the change of
energy-momentum of the particle due to the interaction with the external
field, and the second one describes the precession of the colour spin in
the external nonabelian field.

In the following  we shall confine
our consideration to the one-dimensional case and look at the changes in
the energy of the massive coloured projectile.

It is convenient to rewrite the system of
equations (1) in the laboratory system:
\begin{eqnarray}
m{d v_x \over dt} &=& gQ^a(t) E^a_x(t,x) (1-v_x^2)^{3/2} \nonumber \\
{d {\cal{E}} \over dt} &=& gQ^a(t) E^a_x(t,x) v_x \nonumber \\
{d Q^a \over dt} &=& g f_{abc}(v_x A^b_x(t,x)-
                               A^b_0(t,x)) Q^c(t),
\end{eqnarray}
where the particle is moving along the $x$-axis with an instantoneous
velocity $v_x(t)$, $E^a_x(t,x)$ is a chromoelectric field strength,
${\cal{E}}$ is energy of the particle.
The calculations presented
below are straightforwardly generalizing those by Sturrock [11] in
his discussion of a stochastic plasma heating on a
relativistic non-abelian case.

{\bf{3}}. Let us now discuss the description of the ensemble of the
stochastic chromoelectric field. We assume it is a stationary
ensemble characterized by the lowest order correlators for the group
$SU(N_c)$:
\begin{equation}
\left \{
\begin{array}{lcl}
\langle E^a_x(t,x) \rangle &=& 0 \nonumber \\
\langle E^a_x(t,x) E^b_x(t',x') \rangle &=&
{1 \over 3} {1 \over N^2_c-1} \langle {\bf E}^2 \rangle \delta^{ab}
R(t-t',x-x'),
\end{array}
\right .
\end{equation}
where the brackets denote the averaging over the stochastic ensemble.
The dimensionful quantity $\langle {\bf E}^2 \rangle$ is an average value of
the chromoelectric field squared (nonperturbative condensate). For the
dimensionless correlation function $R$ one has to introduce some
parametrization. The Eucledian counterpart of this correlation
was measured on the lattice [15]. Different
parametrizations of it were
used in phenomenological applications (I.V.Andreev [6], [8]).

Of a crucial importance for the following discussion is the {\it{sign}}
of the condensate $\langle {\bf E}^2 \rangle$. In the (by now) standard
approach to the description of the nonperturbative phenomena in terms of
condensates this quantity is {\it{negative}}. This shows itself, for
example, in the unusual sign of the quadratic Stark effect due to the
stochastic chromoelectric vacuum fields in heavy quarkonia [16].
 At the same time the
sources of chromoelectric fields like a system of heavy charges
considered in the discussions of the QCD analog of the
Landau-Pomeranchuk effect [17] or the instabilities of the quark-gluon
plasma correspond to a {\it{positive}} sign of $\langle {\bf E}^2 \rangle$.
Below it will be shown, that it is precisely this difference in sign
that leads to stochastic decceleration or acceleration of the projectile
particle in the external stochastic chromoelectric field.

We conclude this section by some technical remarks.
In the following we shall work in the lowest (second) order in the
chromoelectric field and suppose that the autocorrelation time of the
stochastic external field is less than a characteristic time scale of a
problem (i.e. $\tau_c < 1$ fm). Another point is that as the third equation in
(3) contains the gauge
potentials it is necessary to express them through the electric field
strength. The most straightforward way to do it is to use a
Fock-Schwinger gauge $x^{\mu} A^a_{\mu}=0$, in which the potentials are
expressed through the field strength by the formula
\begin{equation}
A^a_{\mu} = \int_0^1 d \alpha \alpha x^{\rho} G^a_{\rho \mu}(\alpha x)
\end{equation}
Then the equation for the charge evolution takes the form
\begin{equation}
m{d \over dt} Q^a(t)=g f_{abc} (x+v_x t)
\int_0^1 d \alpha \alpha E^b_x (\alpha t, \alpha x) Q^c (t)
\end{equation}
which we shall use below.

{\bf{4.}} Let us now look at the evolution of the average particle
energy $\langle {\cal{E}} \rangle$. From the Eq. (2) we see that in
order to calculate the right hand side of the second equation in the
second order in $E^a(t,x)$ one has to calculate the changes in
the coordinate, charge and velocity in the first order in the chromoelectric
field. Assuming the unperturbed motion of the form $x_0(t)=v_0 t$,
one gets
\begin{eqnarray}
x(t) &=& v_0 t+{g \over m} Q^a_0 (1-v_0)^{3/2}
\int_0^{t} dt' (t-t') E^a_x(t',v_0 t') \nonumber \\
v_x(t) &=& v_0 + {g Q^a_0 \over m} (1-v_0)^{3/2} \int_0^{t}
dt'E^a_x(v_0 t',t') \nonumber \\
Q^a (t) &=& Q^a_0+{2gv_0 \over m}f_{abc}
\int_0^1 d\alpha \alpha \int_0^{t} dt' t' E^b_x (t',v_0 t') Q^c_0
\end{eqnarray}
where $Q^a_0$ is an initial colour spin.
Substituting the formulas (6)
into Eq.(2) and averaging over the stochastic external field,
we get in the leading order in $E^a$:
$$
\langle {d{\cal{E}} \over dt} \rangle =
{4\pi \alpha_s \langle {\bf E}^2 \rangle \over 3m (N^2_c-1)} \{
(1-v_0^2)^{3/2} (Q_0)^2 \int_0^{\infty} dt' R((t-t'),v_0 (t-t'))
$$
$$
+(1-v_0^2)^{3/2} (Q_0)^2  \int_0^{\infty} dt' v_0 (t-t')
R'((t-t'),v_0 (t-t'))
$$
\begin{equation}
+2 v_0 f_{abc} \delta ^{ab} Q^c_0 \int_0^1d \alpha \alpha \int_0^{\infty}dt' t'
R(t-t',v_0(t-t')) \}
\end{equation}
where the derivative in the second term is taken with respect to a
spatial coordinate
and the infinite limit for the
integration over time follows straightforwardly from the assumption that
the autocorrelation time of the external field is small on the
characteristic time scale of a problem.

{}From the antisymmetry of the structure constants $f_{abc}$ it
immediately follows that the third contribution is equal to zero and we
finally get
\begin{equation}
\langle {d{\cal{E}} \over dt} \rangle =
{4\pi \alpha_s \langle {\bf E}^2 \rangle \over 3m (N^2_c-1)}
(1-v_0^2)^{3/2} (Q_0)^2 \int_0^{\infty} d{\tilde t}
(R({\tilde t},v_0 {\tilde t})+v_0 {\tilde t} R'({\tilde t}
,v_0 {\tilde t}))
\end{equation}

The formula (8) clearly shows that the sign of the energy evolution is
determined by that of an average square of the stochastic chromoelectric
field (condensate) $\langle {\bf E}^2 \rangle$. For the usual nonperturbative
condensate (negative $\langle {\bf E}^2 \rangle$) we have energy losses
(stochastic decceleration), and in the opposite case of a positive
$\langle {\bf E^2} \rangle$ we have the standard stochastic acceleration
well-known in plasma physics.

In order to study the dependence of the energy loss (gain) of the colour
projectile on the autocorrelation time of the stochastic chromoelectric
field $t_c$ we have to introduce a specific parametrization of the
correlation function $R(t-t',x-x')$. Below we shall see that a
choice of this parametrization can dramatically change the resulting
expression for the energy losses.

Let us consider two possible parametrizations of the correlation
function $R$:
\begin{eqnarray}
R_1 &=& exp(-((t-t')^2+(x-x')^2)/t_c^2) \\
R_2 &=& exp(-((t-t')^2-(x-x')^2)/t_c^2)
\end{eqnarray}
Let us stress that the second Lorentz-invariant expression
provides a natural analytical
continuation of one of the parametrizations for the correlation function
studied in Eucledian formalism.
In Eq.(9) we have the correlation function $R$ taken on the
nonperturbed particle trajectory. Thus we get
\begin{eqnarray}
R_1 ({\tilde t}) &=& exp(-(1+v_0^2){\tilde t}^2/t_c^2) \\
R_2 ({\tilde t}) &=& exp(-(1-v_0^2){\tilde t}^2/t_c^2)
\end{eqnarray}

Substituting these parametrization into Eq. (9) we get
\begin{equation}
\langle {d{\cal{E}} \over dt} \rangle _1 =
{2\pi^{3/2} \alpha_s \langle E^2 \rangle \tau_c \over 3m}
({1-v_0^2 \over 1+v_0^2})^{3/2}  {Q_0^2 \over N^2_c-1}
\end{equation}
and
\begin{equation}
\langle {d{\cal{E}} \over dt} \rangle _2 =
{2\pi^{3/2} \alpha_s \langle {\bf E}^2 \rangle \tau_c \over 3m}
{Q_0^2 \over N^2_c-1}
\end{equation}
We can conclude that the usual description of the hadronic medium
(negative $\langle {\bf E}^2 \rangle$) leads to the energy loss of a colour
projectile in the stochastic colour medium thus {\it{softening}} the
energy spectrum of the produced hadrons. At the same time the
stochastic acceleration of heavy quarkonium by stochastic fields in
quark-gluion plasma originating from its instabilities or even
turbulence will {\it harden} this spectrum.

Denoting the initial energy of the projectile by $E_0$
and recalling, that the conserved square of the colour spin  is equal to
[14]
$$
(Q^a_0)^2 = 3 C_A
$$
where $C_A$ is a Casimir operator of the corresponding representation of
the colour qroup (for $SU(N_c)$  $C_V=N_c$, $C_F=(N^2_c-1)/2N_c$ for the
vector and fundamental representations respectively)
 we obtain the final formulas
\begin{equation}
\langle {d{\cal{E}} \over dt} \rangle _1=
{2\pi^{3/2} \alpha_s \langle {\bf E}^2 \rangle \tau_c \over m}
({2 E_0^2 \over m^2}-1)^{-3/2} {C_A \over N^2_c-1}
\end{equation}
and
\begin{equation}
\langle {d{\cal{E}} \over dt} \rangle _2=
{2\pi^{3/2} \alpha_s \langle {\bf E}^2 \rangle \tau_c \over m}
 {C_A \over N^2_c-1}
\end{equation}
We see that the energy dependence of the energy losses rate has a
dramatic dependence on the form of the correlation function.

For the intuitively appealing (although Lorentz-noninvariant) choice
$R_1$
the rate of the energy losses (gains) decays (at sufficiently high
energies of the incident particle $E_0$) as $E_0^{-3}$ and thus is
numerically negligible.

The second Lorentz-invariant choice $R_2$ leads to an energy independent
energy loss (gain) rate and thus can be quantitatively important.

Let us estimate the corresponding energy losses for the charmed quark
(fundamental representation of $SU(3)$) and the $J/\Psi$ meson in the
octet state (adjoint representation of $SU(3)$) in the case of a constant
energy change rate. Using
$$
\langle \alpha _s {\bf E}^2 \rangle =
- {1 \over 4} \langle \alpha_s G^2 \rangle,
$$
$m_c=1.5 \mbox { GeV }$, $\langle \alpha_s G^2 \rangle = 0.042
\mbox { GeV}^4$ and $t_c = 0.3 \mbox { fm }$ we get
$$
\langle {d{\cal{E}} \over dt} \rangle _q =
98 {\mbox MeV \over \mbox fm}
$$
and
$$
\langle {d{\cal{E}} \over dt} \rangle _{J/\Psi} =
110 {\mbox MeV \over \mbox fm}
$$

Taking into account the additional numreical
uncertainities (the value of $t_c$, etc) we see, that
the nonperturbative decceleration of the described type can be
around 10-20 percent with respect to a naively expected stringy
 energy loss in the hadronic medium.

Concluding this section we note, that
the formulas (15) and (16)
illustrate (within logarithmic accuracy) the borderline energy
dependences following from the simplest choice of a correlation
function characterizing the ensemble of the stochastic chromoelectric
external fields.

{\bf{5.}} Let us briefly summarize the results.
Considering the quasiclassical equations of motion for a heavy coloured
particle  propagating in the external
stochastic chromoelectric field we have calculated the energy evolution
rate. The calculations are generalizing to a non-abelian
relativistic case those explaining the stochastic
heating phenomenon in plasma physics.
We have
shown that the cases of a projectile decceleration and acceleration are
distinguished by the sign of an chromoelectric field condensate. The
negative sign corresponds to stochastic decceleration (energy loss) and
the positive one to stochastic acceleration (energy gain). The energy
dependence of the rate of the energy loss turns out to be very sencitive
to the form of the correlation function of the stochastic
external chromoelectric field.

The first possibility (negative $\langle E^2 \rangle$) is believed to be
realized in the nonperturbative
colour medium leading for example to an unusual sign of a level shift of
heavy quarkonium due to a quadratic Stark effect [16]. The experimental
prediction will be the
corresponding softening of the energy spectrum of the produced
hadrons.

The most natural situation when the second possibility (positive
$\langle E^2 \rangle$) is realized is for hadron production  in the
quark-gluon plasma in the presence of an instability (turbulence) or a
propagation of coloured projectile originating from the hard stage
of a collision through it.
The experimental prediction is the hardening of the energy spectrum of the
produced hadrons. This phenomenon is  analogeous to the stochastic
acceleration of charged particles in the cosmic plasma responcible
 for the formation of the
energy spectrum of cosmic rays [12].

\begin{center}
{\it{Acknowledgements}}
\end{center}

I am grateful to Prof. H.Satz for kind hospitalty in the Bielefeld
University. The work was partially supported by the Russian Fund ofr
fundamental Research, grant 93-02-3815.

I would like to thank R.Baier, D.Kharzeev, H.Satz and G.Zinovjev for reading
the manuscript and very useful discussions.

\begin{center}
{\it{References}}
\end{center}

1. H.Fritzch {\it{Phys.Lett}} {\bf{B67}} (1977), 217;

   M.Gl${\ddot u}$ck, E.Reya {\it{Phys.Lett}} {\bf{B79}} (1978), 453;

   E.L.Berger, D.Jones {\it{Phys. Rev}} {\bf{D23}} (1981), 1521;

2. M.A.Shifman, A.I.Vainstein and V.I.Zakharov
{\it{Nucl.Phys.}} {\bf{B147}} (1979), 385;

3. T.Matsui, H.Satz {\it{Phys.Lett}} {\bf{B178}} (1986), 416;

4. G.Piller, J. Mutzbauer and W.Weise {\it{Nucl.Phys.}} {\bf{A560}}
(1993), 437;

5. D.Kharzeev, H.Satz {\it{Z.Phys.}} {\bf{C60}} (1993), 389;

6. R.Baier, R. R${\ddot u}$ckl {\it{Z.Phys.}} {\bf{C19}} (1982), 251;

7. P.Olesen {\it{Nucl.Phys}} {\bf{B200}} (1982), 381;

   S.M.Apenko, D.A.Kirzhnitz and Yu.E.Lozovik {\it{JETP Lett.}} {\bf{36}}
(1982), 213;

   I.V.Andreev {\it{JETP Lett.}} {\bf{41}}(1985), 592;

   H.G.Dosch, Yu.A.Simonov {\it{Phys.Lett.}} {\bf{B205}}(1988), 339;

8. H.G.Dosch, U.Marquard {\it{Nucl.Phys.}} {\bf{A560}} (1993), 333;

   Yu.A.Simonov, Preprint ITEP 84-93;

9. D.Gromes {\it{Phys.Lett.}} {\bf{B115}} (1982), 482;

10. H.G.Dosch, E.Ferreira and A.Kramer {\it{Phys.Lett.}} {\bf{B289}}
(1992), 153;

11. P.A.Sturrock {\it{Phys.Rev.}} {\bf{141}} (1966), 186;

12. V.N.Tsytovich, "An Introduction to the Theory of Plasma Turbulence",
Pergamon Press, 1972;

13. S.K.Wong {\it{Nuovo Cim.}} {\bf{LXVA}}, No. 4 (1970), 689;

14. U.Heinz {\it{Ann.Phys.}} {\bf{161}} (1985), 48;

15. A.Du Giacomo, H. Panagopoulos {\it{Phys.Lett.}} {\bf{B285}} (1992),
183;

16. M.B.Voloshin {\it{Nucl.Phys.}} {\bf{B154}} (1979), 365;

    H. Leutwyler {\it{Phys. Lett.}}{\bf{B98}} (1981), 447;

17. M.Gyulassy, X.-N.Wang, Preprint LBL-32682 (1993).

\end{document}